\documentclass[prb,twocolumn,showpacs,preprintnumbers,amsmath,amssymb]{revtex4}

\usepackage[]{graphicx}			
\usepackage{bm}				
\usepackage{times}				

\newcommand{\za}{\ensuremath{\alpha}}

\newcommand{\ohm}{\ensuremath{\Omega}}

\begin{document}

%
%

\title{Current-driven microwave oscillations in current perpendicular-to-plane spin-valve nanopillars}
\author{Q. Mistral}
\email{quentin.mistral@ief.u-psud.fr}
\author{Joo-Von Kim}
\author{T. Devolder}
\author{P. Crozat}
\author{C. Chappert} 
\affiliation{Institut d'Electronique Fondamentale, UMR CNRS 8622, Universit\'{e} Paris-Sud, 91405 Orsay cedex, France} 
\author{J. A. Katine}
\author{M. J. Carey}
\affiliation{Hitachi Global Storage Technologies, San Jose Research Center, 650 Harry Road, San Jose, California 95120, USA}
\author{K. Ito}
\affiliation{Hitachi Cambridge Laboratory, Madingley Road, Cambridge CB3 0HE, United Kingdom}
\date{\today}

%
%
\begin{abstract}

We study the current and temperature dependences of the microwave voltage emission of spin-valve nanopillars subjected to an in-plane magnetic field and a perpendicular-to-plane current. Despite the complex multilayer geometry, clear microwave emission is observed for current densities in the interval $9 \times 10^{7}$ to $13 \times 10^{7}$~A cm$^{-2}$. The emission frequency stays near 12 GHz when $I<I_{red} =11.2 \times 10^{7}$~A cm$^{-2}$, then redshifts with a slope gradually reaching -350 MHz/mA for $16 \times 10^{7}$~A cm$^{-2}$. The linewidth narrows exponentially to 3.8 MHz at 150K for $I < I_{red}$, then broadens again as the emitted voltage redshifts. The temperature dependence of the linewidth exhibits a curvature change around the linewidth minimum.

\end{abstract}


\maketitle

%
%

	Since the seminal proposal of L. Berger~\cite{Berger:PRB:1996} and J. Slonczewski~\cite{Slonczewski:JMMM:1996} predicting that a spin-polarized current would exert a torque onto the magnetization, extensive experimental~\cite{Myers:Science:2000},\cite{Tsoi:Nature:2000} and theoretical~\cite{Slonczewski:JMMM:1999},\cite{Stiles:PRB:2002} work has been directed toward understanding the underlying physics in view of future applications. The primary effect on these length scales is the transfer of spin angular momentum between the spin-polarized current and the magnetization. This results in a spin torque (ST) that may either lead to magnetization reversal or to magnetization oscillations~\cite{Kiselev:Nature:2003}.
	In these stationary dynamical states, magnetization oscillates along a fixed trajectory, because there is a compensation over one period between the energy supplied by the ST and the  energy lost by the magnetization as a result of damping. Because these systems have magnetoresistance, the applied current and this stationary magnetization oscillations can result in emission peaks in the microwave spectrum. This holds promise for the design of integrated compact-sized, tunable and agile microwave oscillators.

ST oscillators have now been implemented in different geometries,~\cite{Sankey:PRB:2006, Rippard:PRL:2004} materials,~\cite{Covington:PRB:2004} and magnetic configurations~\cite{Kiselev:PRL:2004, Rippard:PRB:2004,Kiselev:PRB:2005}. While the reported frequency linewidths can be very narrow, there is a significant scatter (250~MHz~\cite{Kiselev:PRL:2004} to 2~MHz~\cite{Rippard:PRB:2004}).
	In this article, we present a study of the current and temperature dependence of the microwave oscillations induced by ST in a nanopillar.
Our samples are complex multilayers similar to standard read-head stacks. There is no obvious polarizing layers, which is a major diffrence with the much studied pseudo spin valve stack (e.g. Co/Cu/Co). Despite this, we observe peaks in the emission spectra with linewidths as narrow as 3.2 MHz, which is comparable to linewidths obtained for point-contact geometries.~\cite{Rippard:PRB:2004} This is remarkable because self-localization of spin-wave excitations~\cite{Slavin:PRL:2005} is usally cited as the reason for which point-contact linewidths are consistently smaller than their counterparts in nanopillars. We show here that comparable linewidths can be obtained for nanopillars comprising complex multilayer structures.

The samples consist of sputtered spin-valve films patterned with e-beam lithography, similar to those studied in Ref.~\onlinecite{Lacour:APL:2004}. The film has the following material composition: Ta (5)/Cu (20)/Ta (3)/Cu (20)/Ta (2.5)/PtMn (17.5)/CoFe (1.8)/Ru (0.8)/CoFe (2)/Cu (3.5)/CoFe (1)/NiFe (1.8)/Cr (5)/Au (150), where the figures in parentheses indicate the film thickness in nanometers. The free layer is the composite CoFe (1)/NiFe (1.8) bilayer.  This structure is completely etched in a rectangular section of 50 $\times$ 100 nm$^2$ and isolated before electrical contact is made. The top electrode is contacted to a coplanar wave guide and the bottom electrode is shorted to the ground. In this structure, 1~mA corresponds to $2 \times 10^7$~A.cm$^{-2}$.

\begin{figure}
\includegraphics[width=6cm]{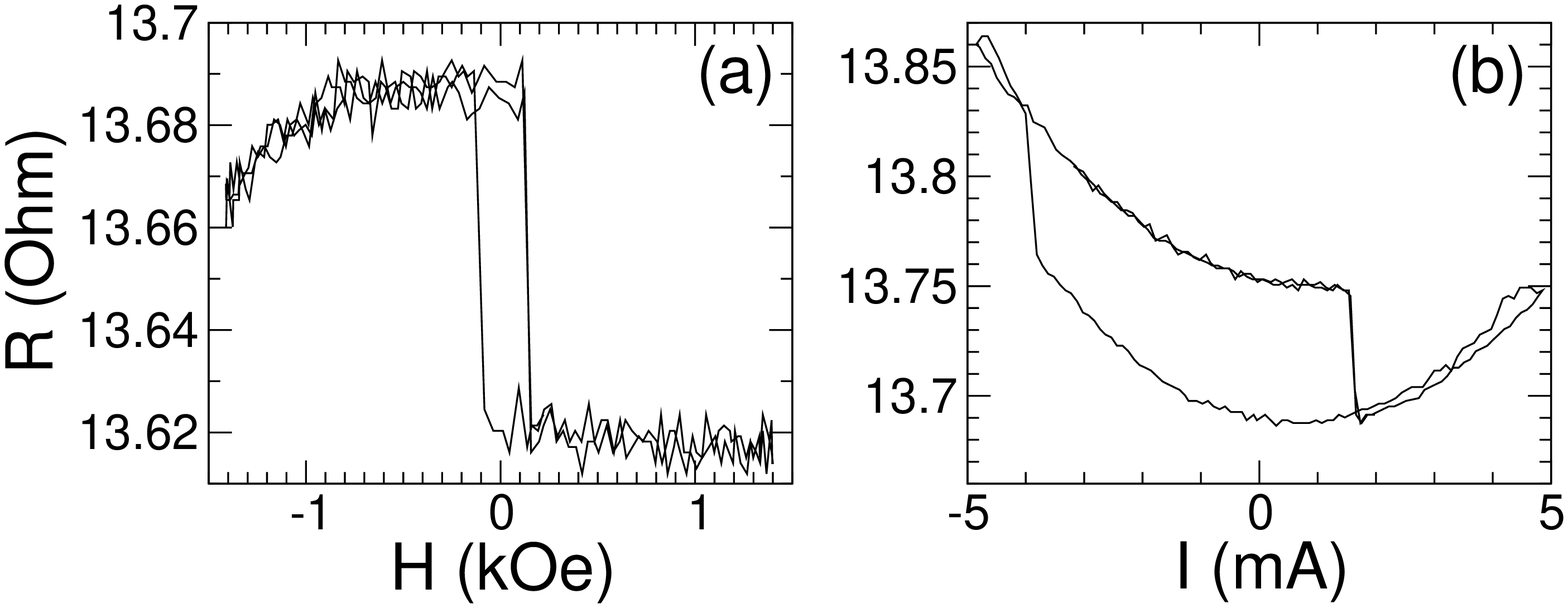}
\caption{\label{fig:GMR}Device magnetoresistance as a function of (a) applied magnetic field $H$ along free layer easy axis (DC current set to 80~$\mu$A, 80~kHz) and (b) applied electric current $I$ (applied magnetic field set to 0 Oe).}
\end{figure}

Static measurements of the device reveal a 60 m$\ohm$  magnetoresistance for both resistance versus current $I$ and resistance versus magnetic field $H$ curves (Fig.~\ref{fig:GMR}).
From the measurement with $H$ applied along the hard axis direction, we deduce the total uniaxial anisotropy of the free layer to be $13 \pm 1$ mT and a bias field of 3 mT is observed in the hysteresis loop, which we attribute to stray dipolar fields from the other magnetic layers in the stack. From previous studies,~\cite{Lacour:APL:2004} the saturation magnetization $\mu_0 M_s$ is deduced to be 0.85~T with a Gilbert damping constant of $\za = 0.02$. The critical currents  are 1.6 mA for the antiparallel (AP) to parallel (P) transition and -3.93 mA for the P to AP transition. The dc resistance is 13.6 $\ohm$ at room temperature. In this experiment, the nanopillar is biased with a constant $-1.3$ kOe magnetic field along the easy axis of the free layer, which favors the antiparallel state. 

In order to identify the microwave-active region, we first look for anomalies in the resistance versus current curve at fixed applied field. Once an anomaly has been detected, we acquire the power spectral density of the microwave emission using a spectrum analyzer after with a 43 dB amplification, with a bandwidth of 0.1 GHz to 26 GHz. More details about the setup can be found elsewhere.~\cite{Devolder:PRB:2005, Mistral:MatSciEngB:2005}.
The resolution bandwidth (RBW) of our spectrum analyzer was set to 2 MHz. The convolution function of the RBW filter has been calibrated by analysing the system's response to spectrally pure synthezised signals. This calibration has been used to correct experimental linewidth.
We estimate the maximum uncertainty, which are typically 10 MHz for the emission frequency, 1.7 MHz for the linewidths and $0.5 \times 10^{-7}$ nV$^{2}$ for  the power. We point out that the parameter most susceptible to errors is the peak amplitude, and not the linewidth.
We extracted the signal characteristics from Lorentzian and Gaussian fits to the spectral lines, from which we obtained total power (area under the curve), frequency, and the full-width-at-half-maximum (FWHM) linewidth. Measurements of the power emission spectra were made at three temperatures: 150 K, 185 K and 225 K. 

\begin{figure}
\includegraphics[width=7cm]{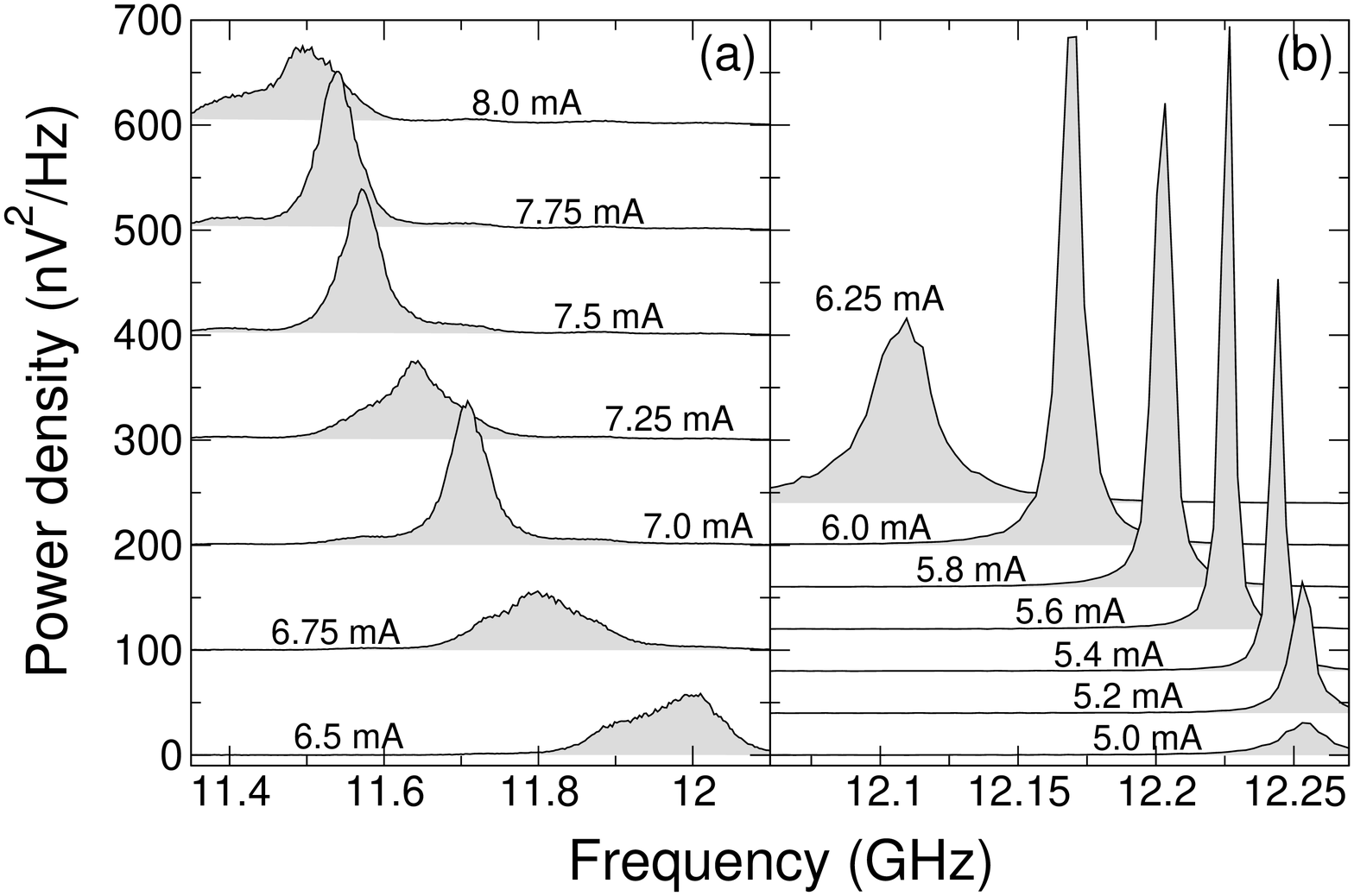}
\caption{\label{fig:Raw_Data}Current-induced magnetoresistance oscillations of the spin-valve nanopillar, for the (a) high current and (b) low current regimes at T=150 K. A vertical offset is applied to the curves for visual clarity. }
\end{figure}

\begin{figure}
\includegraphics[width=5.5cm]{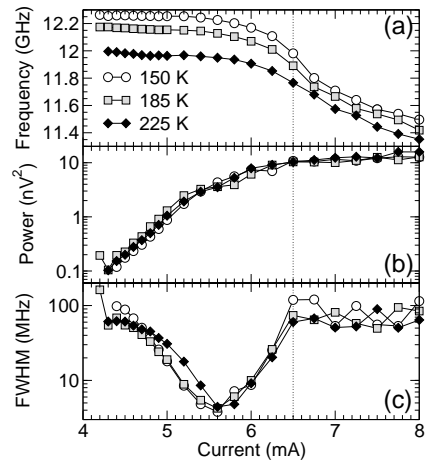}
\caption{\label{fig:PvsI}(a) Frequency, (b) power and (c) linewidth (FWHM) of the fundamental mode as a function of applied current. The dashed line at 6.5 mA indicates the apparition of a different mode regime.}
\end{figure}

In Fig.~\ref{fig:Raw_Data}, we present the power spectrum density of the signal emitted by our nanopillar when subjected to a dc electric current.
For currents below $I=$ 4.2 mA, we observe no high-frequency signal above the noise floor.
Measurements of two successive series of peaks are observed. The first series occurs in the interval 4.2 mA $\lesssim I < $ 6.5 mA, in which a well-defined Lorentzian lineshape is observed. The second series is seen for 6.5 mA $\lesssim I < $ 8.0 mA and is characterized by largely deformed lineshapes with significantly broader linewidths than the first series. Above 8.0 mA (measurements were made up to 10 mA), large electrical resonances were seen which drown out the magnetoresistance oscillations. A second order harmonic is also observed (not shown) at twice the frequency of the peaks shown in Fig.~\ref{fig:Raw_Data}.
Quantitative data are gathered in Fig.~\ref{fig:PvsI}, where we report the frequencies, powers, and linewidths of the main resonance as a function of applied current for the three studied temperatures.
%
First, we note that the oscillation frequencies decrease with increasing current in both regimes. This trend is consistent for the three temperatures considered, and has also been observed in other systems for which the applied magnetic field is in the film plane. The decrease in frequency for the low-current regime is relatively slow in comparison with the high-current regime. Second, we observe an exponential increase in the total power emitted (over the entire frequency spectrum measured) as a function for current in the low-current regime, which levels off as the high-current regime is attained (Fig.~\ref{fig:PvsI}b). We point out that a similar trend is obtained if we restrict the integration to the oscillation peak, which indicates that little power is pumped out of the main resonance and its second harmonic.

The variation of the oscillation frequencies as a function of applied current exhibits a more complex behavior than that observed in point-contact geometries. According to spin-wave theory,~\cite{Rezende:PRL:2005,Slavin:IEEE:2005} the frequency shift is a self-induced nonlinear effect that scales with the square of the oscillation amplitude. For an in-plane quasi-uniform precessional mode (i.e. one that closely resembles the ferromagnetic resonance mode), one finds a current dependence of the mode frequency of the form (Eq.  43 in Ref.~\onlinecite{Slavin:IEEE:2005}) where $I_{\text{th}}$ is the threshold current at which self-sustained magnetic oscillations are made possible. While this equation finds some quantitative agreement for point-contact geometries, it does not describe the slow frequency variation and concave curvature of our experimental $f(I)$ curves in the low-current regime. A better qualitative agreement is achieved in the high-current regime, but the experimental lineshapes are distorted, and one may equally speculate that these broad peaks reveal the excitation of several magnon modes with close frequencies. 

We surmize that the oscillations observed are due to in-plane precession modes. Not only is this consistent with the frequency redshift predicted from spin-wave theory~\cite{Rezende:PRL:2005,Slavin:IEEE:2005} and macrospin simulations,~\cite{Xiao:PRB:2005} the dynamic magnetoresistance measured also lends credence to this hypothesis. After amplification, the highest power density we measured is 24 dB above the noise floor (-103 dBm) for 5.6~mA. If we consider the spin transfer oscillator as a voltage generator, we obtain an estimate of the amplitude of the dynamic magnetoresistance to be $6 \times 10^{-4}$ \% of total magneto resistance. For out-of-plane precession, one would expect this amplitude to be of the same order of magnitude as the static magnetoresistance measured (i.e. $\sim$ 60 m$\ohm$), based on the magnetization trajectories expected for such a mode of oscillation.~\cite{Xiao:PRB:2005}

The central result of this article is shown shown in Fig.~\ref{fig:PvsI}c, where the fundamental mode linewidth is shown as a function of current. We observe a relatively broad linewidth of the order of 100 MHz just above the noise threshold at low currents. The linewidth then narrows quasi-exponentially as the current is increased and reaches a minimum at about $I = $ 5.6 mA, where for $T = $ 150 K we obtain a linewidth of 3.2 MHz (after deconvolution with the RBW filter) which is comparable to the linewidths observed in point-contact geometries~\cite{Rippard:PRB:2004}. This result was obtained on one sample which exhibits high signal. On a similarly designed device (50~$\times$~100~nm$^2$), a minimium of 18 MHz was obtained with less amplitude. As the current is increased again past this minimum, an exponential broadening in the linewidth is seen. This trend is consistent for the three temperatures considered. We point out that this minimum occurs well within the low-current regime, and that the broadening of linewidth with current is not due solely to a deformation of the lineshape. Indeed, the linewidth plateau in the high-current regime can be attributed to such deformations.
\begin{figure}
\includegraphics[width=6.5cm]{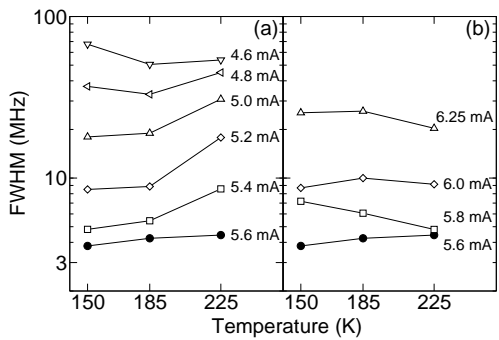}
\caption{\label{fig:WvsT}Temperature dependence of the linewidth for increasing applied currents (a) towards and (b) away from the linewidth minimum.}
\end{figure}

The initial narrowing in linewidth can be understood from stochastic arguments. Just above the threshold, the dynamical state is dominated by amplitude and phase fluctuations due to thermal noise (e.g. thermal magnons), which results in the broad spectral lines observed. As the current is increased, the amplitude of precession increases and, as a consequence, the trajectories become more immune to amplitude fluctuations (which scale with temperature and not amplitude). In the limit of large amplitude motion in which only phase noise  contributes, the spectral linewidth can be estimated from stochastic  spin-wave theory~\cite{Kim:PRB:2006}. The excited spin-wave mode is assumed to  be buffeted by random forces as a result of thermal noise, and it is  shown that the resulting linewidth is inversely proportional to the  spin-wave intensity. As such, the narrow linewidths observed is a  consequence of a well-defined large amplitude spin wave mode. We  conjecture that the stability of this mode might be a result of  strong quantization effects in our rectangular elements. 

For the linewidth broadening past the linewidth minimum, we speculate that this is due to higher-order nonlinear effects such as the generation of closely spaced modes in frequency space. To elucidate this behavior further, we present the temperature dependence of the linewidth around the linewidth minimum in Fig.~\ref{fig:WvsT}. In Fig.~\ref{fig:WvsT}a, the linewidth variation is shown for increasing currents that lead to linewidth narrowing, while in Fig.~\ref{fig:WvsT}b, the variation is shown for increasing currents leading to linewidth broadening. There are some noticeable differences between the two regimes. First, in Fig.~\ref{fig:WvsT}a, we observe that the linewidths broadening with temperature with a convex curvature, which can be expected from the stochastic arguments outlined above.  As the current is increased past the linewidth minimum, however, the linewidths decrease as a function of temperature with a concave curvature, which indicates another physical process at work.


In summary, we present evidence of large current-driven magnetoresistance oscillations in spin-valve nanopillars, with linewidths as narrow as 3.2 MHz for a fundamental peak at 12.23 GHz. The current and temperature dependence of these oscillations reveal a more complex behavior than that observed in point-contact and other pseudo spin-valve geometries. 

Q. Mistral wishes to acknowledge partial support from ST Microelectronics under a joint CNRS PhD grant.


\end{document}